\documentclass[twocolumn,showpacs,preprintnumbers,amsmath,amssymb,aps,pre]{revtex4}
\usepackage{graphicx}
\begin{document}

\title{
Structural Transitions, Melting, and Intermediate Phases for Stripe and 
Clump Forming Systems 
} 
\author{
C. J. Olson Reichhardt, C. Reichhardt, and A.R. Bishop} 
\affiliation{
Theoretical Division,
Los Alamos National Laboratory, Los Alamos, New Mexico 87545, USA } 

\date{\today}
\begin{abstract}
We numerically examine the properties of 
a two-dimensional system of particles which have competing long range repulsive 
and short range attractive interactions 
as a function of density and temperature. 
For increasing density, there are well defined transitions 
between a low density 
clump phase, an intermediate stripe
phase, an anticlump phase, 
and a high density uniform phase. 
To characterize the transitions between these phases 
we propose several measures which take into
account the different length scales in the system. 
For increasing temperature, we find 
an intermediate phase that is liquid-like on the short length scale of
interparticle spacing but solid-like on the larger length scale of the
clump, stripe, or anticlump pattern.
This intermediate phase persists over the widest temperature range in the
stripe phase when the local particle lattice
within an individual stripe melts well below
the temperature at which the entire stripe structure breaks down, and
is characterized by
intra-stripe diffusion of particles
without inter-stripe diffusion. 
This is
followed at higher temperatures by the onset of inter-stripe diffusion in
an anisotropic diffusion phase, and 
then by breakup of the stripe
structure.
We identify the transitions between these regimes 
through diffusion, specific heat,
and energy fluctuation measurements, 
and find that within the intra-stripe liquid regime, the excess entropy
goes into disordering the particle arrangements within the stripe rather
than affecting the stripe structure itself.
The clump and anticlump phases
also show multiple temperature-induced 
diffusive regimes
which are not as pronounced as those of the stripe
phase. 
\end{abstract}
\maketitle

\vskip2pc
There has been growing interest recently in understanding 
how collections of particles with competing interactions order
or disorder under different conditions. 
A general understanding of 
such
behaviors would have a profound 
impact on our ability to control self-organization and to tailor the  
functionality in systems of this class \cite{Andelman,Pell,Glaser},  
such as collections of particles
with completing interactions composed of a long-range repulsion 
and a short range attraction 
\cite{Bishop,Martin,Olson,Peeters,Yu,Bishop3,Colloid}. 
Specific systems where this type of interaction may be realized include 
charge ordering in cuprate superconductors \cite{Saxena,Stroud,Kabanov},
magnetic systems \cite{Seul,Singer}, 
bubble, stripe and clump phases in two-dimensional electron 
systems \cite{Kivelson,Lilly},  
and dense nuclear matter which forms pasta phases \cite{Ra}. 
There are also a number of soft matter systems which can 
be effectively modeled
as assemblies of
particles with a finite range repulsive interaction that competes with
a shorter range
attractive interaction
\cite{Babic,Zapperi,Davidov}.    
Recent simulations
of colloidal systems with this type of interaction 
\cite{Zapperi} 
generated many of the same types of phases 
observed in models with longer range repulsive 
interactions \cite{Bishop3},
indicating that many of the features found in systems with 
long range repulsion
and short range attraction 
will be relevant for other systems with 
competing interactions as long the repulsive 
interaction
is longer range than the attractive 
interaction.
Clump, stripe, and bubble phases have also been shown to occur in
systems where the particle-particle interactions  
are strictly repulsive 
provided that there are two length scales present in the repulsive 
interaction \cite{Pell,Glaser}.  

For systems with long-range repulsive 
interactions and short range attractive
interactions, the onset of the different clump, stripe, 
and bubble phases can be controlled 
by directly varying the relative strength of the repulsive and attractive
interactions at fixed density, 
or by holding the interactions fixed and varying the 
density of the system. In the clump phase  
where groups of particles form individual islands,
the clumps organize into a crystal pattern at larger length scales
due to the long range repulsive interaction between adjacent clumps
\cite{Martin,Peeters,Bishop3,Colloid,Zapperi}. 
Other phases include stripe or labyrinth structures and anticlump phases 
in which 
an ordered array of voids forms, along with a uniform phase that appears at 
very high densities.
It is not known
whether the transitions between the different phases are well defined, 
whether there are additional subphases, 
or what is the nature of the disordering or melting transitions of the phases
in the presence of thermal fluctuations.

Our previous work on 
the long-range competing interaction system
employed fixed density samples 
in which the relative attractive strength
of the potential was varied \cite{Martin}.
We showed that for narrow regions of parameter space, the system
forms a labyrinth or disordered stripe phase.
Ordered stripes can be generated when the system is 
driven in the presence of quenched disorder; in this case,
the stripes are aligned in the direction of the applied drive \cite{Martin}. 
In the absence of quenched disorder, 
although the particles move under 
an applied uniform drive they do not experience the local 
shearing forces which are necessary to induce the plastic deformations
required to align the stripe pattern.
Simulations of 
a stripe-forming colloidal system with intermediate-range 
repulsive forces indicate 
that if a non-uniform drive is applied so as to create a shear in the system,
then plastic deformation can arise and the 
stripes 
align
in the direction of the shear \cite{Zapperi}. 
For systems with long range repulsive 
interactions,
it is generally 
difficult to 
reach a completely ordered
ground state due to the inherent heterogeneity of the system.
For example, 
if one patch of the system has a higher particle density than 
neighboring patches, it is difficult
for an individual particle to escape from the overly dense patch 
due to the short range attractive forces, and it is also difficult
for a particle to move into an underdense patch of the system due to the
repulsive long range interactions 
\cite{Olson,Bishop3}. 
Colloidal simulations with 
shorter range repulsive interactions also produced similar results 
and showed the formation of 
partially ordered states \cite{Zapperi}. 
 
In this work we simulate the different types of states 
that arise as a function of particle density
in a system with long range repulsive and short range
attractive interactions.
We introduce several new diagnostics to characterize 
these states and the transitions between them.
We also examine how the states disorder 
as a function of temperature. 
The stripe states are particularly useful in understanding 
the thermal disordering process
since diffusion along the stripes 
can be distinguished readily from diffusion between the stripes.
We show that the stripe phase exhibits four temperature induced regimes, 
including a low temperature frozen phase, 
a phase where the particles diffuse along each stripe but where there is no
particle transport between stripes,
a regime where there is anisotropic diffusion 
but the stripe structure persists, and a high temperature phase where the
stripe structure is destroyed. 
The onsets of these different phases produce signatures in
the energy measures and energy fluctuations.
We argue that since the stripe state
can show an intra-stripe melting transition, the stripe structure
may remain stable up to higher temperatures than the other 
possible structures,
since the excess fluctuations in the stripe state 
can first melt the particles within each stripe rather than the stripe
structure itself.
We also show that similar 
melting behaviors occur in the clump and anticlump states.  

\section{Simulation} 
We consider a two-dimensional Langevin simulation of a system
of size $L_x \times L_y$ with 
periodic boundary conditions in the $x$ and $y$-directions.
In order to potentially accommodate a triangular lattice, we
take $L_y=L$ and $L_x=1.097L$.
Within the system we place $N$ interacting particles at a density 
$\rho = N/(L_xL_y)$.
In previous simulations we held the system size fixed and 
either varied the density by changing $N$ 
or varied the relative strength of the attractive and repulsive
particle interactions.  Here we hold the particle-particle 
interaction strength fixed and fix $N=380$ but vary $L$ in order
to alter the density.  
The initial particle positions are obtained using simulated
annealing.
The dynamics of a single particle $i$ obeys the following classical equation of 
motion: 
\begin{equation}  
\eta \frac{d {\bf R}_{i}}{dt} = -\sum^{N}_{j\ne i}{\bf \nabla}V(R_{ij}) +  {\bf F}^{T}_{i} .
\end{equation} 
The particle-particle interaction has the following form:
\begin{equation} 
V(R_{ij}) =  \frac{1}{R_{ij}} -Be^{-{\kappa} R_{ij}}  
\end{equation} 
where $R_{ij}=|{\bf R}_{i}-{\bf R}_{j}|$ is the distance between
particles located at ${\bf R}_{i}$ and ${\bf R}_{j}$.
The first term is a Coulomb repulsion,
meaning that for very short distances the 
interaction is repulsive and the particles will not collapse to a point. 
The magnitude of the fluctuations experienced by the particles 
during the course of the simulation is such that
$R_{ij}$ never drops below the minimum value $R_{ij}=0.1$.
Due to the long range nature of the Coulomb interaction, it is not possible
to cut off the interaction range, so we use a Lekner real-space summation
method to compute the 
long range term \cite{Lekner,Niels,Mazars}.
The second term is a short range attraction 
of Yukawa form with an inverse screening length of 
$\kappa=1.0$ and with $B=2.0$. 
The last term in Eq.~1 is the random fluctuation force
from finite temperature which has the following temporal correlations: 
$\langle F^{T}(t)\rangle = 0$ 
and $\langle F^{T}_{i}(t)F_j^T(t^{\prime})\rangle = 
2\eta k_{B}T\delta_{ij}\delta(t - t^{\prime})$.   

\begin{figure*}
\includegraphics[width=6.2in]{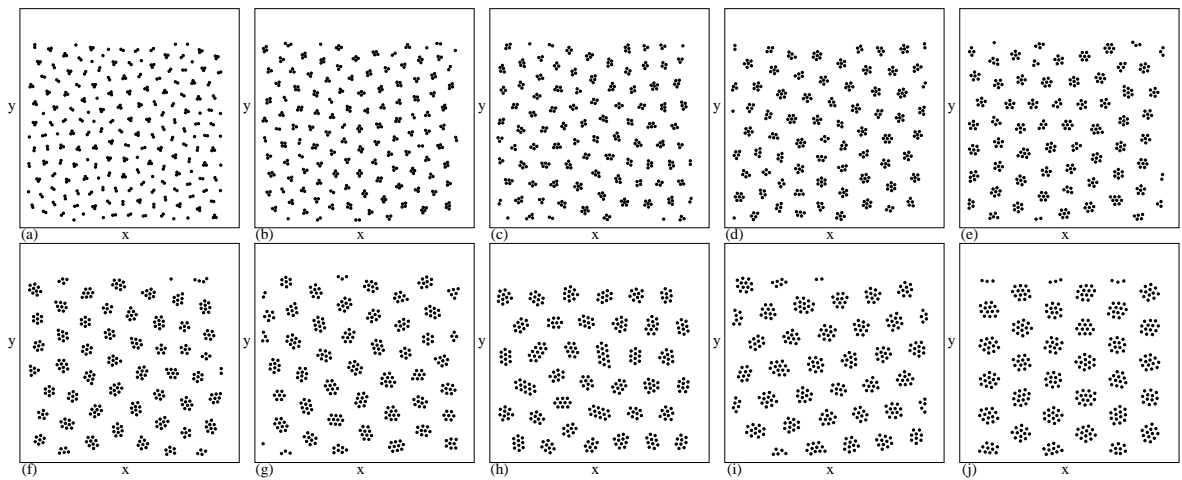}
\caption{
Particle locations in the entire sample after simulated annealing 
for different densities in the clump regime. 
The scale of the x and y axes changes in each panel.
(a) $\rho=0.04$, (b) $\rho=0.06$, (c) $\rho=0.08$, (d) $\rho=0.10$,
(e) $\rho=0.12$, (f) $\rho=0.14$, (g) $\rho=0.16$, (h) $\rho=0.18$,
(i) $\rho=0.20$, and (j) $\rho=0.22$.
The number of particles per clump increases as the density increases. 
At the lowest densities the overall clump structure is disordered; 
however, for $\rho \geq 0.2$ the clumps form a triangular lattice.  
}
\end{figure*}

\section{Phases as a function of Density} 

\subsection{Clump Phases} 
In Fig.~1(a-j) we plot the particle positions after simulated
annealing in the low density regime 
$\rho=0.04$ to $\rho=0.22$.
At the lowest density $\rho=0.04$ in Fig.~1(a), 
a mixture of clumps appears where there are two to three particles in
each clump along with a smaller number of individual particles.
The number of particles per clump increases with increasing $\rho$ so
that at $\rho=0.10$, for example, there are six particles per clump on 
average.
The decrease in the number of particles per clump 
with decreasing density that we find
at the lowest densities suggests that for even lower density
than we have considered, the limit of
only one particle per clump would occur and a triangular lattice would
form.
The system failed to form an ordered ground state at the lowest densities
we considered
even for a very slow simulated annealing procedure.  This results from the
very strong clumping tendency induced by the short range attraction combined
with the relatively large inter-clump distances at these densities.  Particles
are strongly attracted into clumps, but the difference in energy at larger
scales between having a clump with $n$ particles or $n+1$ particles is too
small to permit the particles that have already fallen into a clump to 
escape from the strong short-range attraction of the clump 
in order to form clumps of uniform size.
The resulting polydispersity in clump size prevents the clumps from ordering
into a larger-scale lattice.

Since the long-range Coulomb repulsion favors a triangular ordering
of the clumps, the problem of trying
to order a mixture of clumps of slightly different sizes
has similarities to the filling of magnetic flux quanta at submatching fields
on a triangular pinning array in a superconductor \cite{Jensen}.
At matching fillings where the number of flux quanta is an integer 
multiple $f$ of the number 
of holes, such as at $f = 2$, $f=3$, or $f=n$, with $n$ integer, a triangular
flux lattice forms; however, at non-integer fillings, 
most of the configurations are highly degenerate 
and disordered states appear, such as at $f=n + 1/2$.
At other fields such as $f=n/3$, where an ordered tiling of the
triangular lattice is possible, 
the ground state has several degenerate orientations, 
so any ordered state that forms is very likely to contain grain boundaries.
Analogously, in our system 
a possible partial ordering could be expected for a low density filling 
where 1/3 of the clumps capture $n + 1$ particles and
the remaining $2/3$ of the clumps contain $n$ particles.
In superconducting network systems of the type considered in 
Ref.~\cite{Jensen}, the triangular ordering favored by the repulsive
vortex-vortex interactions is strongly enhanced by the 
ordered substrate. 
In our system, the clumps move over a substrate-free continuum,
which adds even more degeneracy to the system and makes it very 
difficult to observe ordered states.   
Even though there is a difference of only one or two particles in the
number of particles in neighboring clumps, 
the polydispersity of clump sizes is largest at the lowest densities
we consider, where the charge ratio between neighboring clumps can be
as large as $1:2$ or $1:3$.
It is known that large polydispersity in the particle interactions
for two dimensional systems can lead to 
the formation of a disordered state \cite{Reza}.    

\begin{figure*}
\includegraphics[width=6.2in]{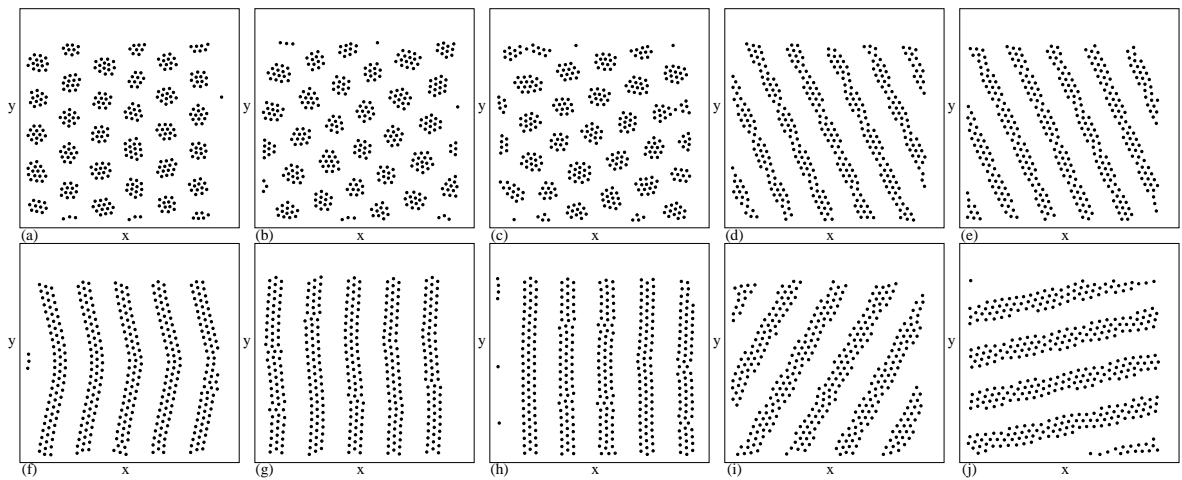}
\caption{
Particle locations in the entire sample after simulated annealing 
for different densities in the clump and stripe regimes. 
The scale of the x and y axes is not fixed in all panels.  
(a) 
$\rho= 0.24$, (b) $\rho=0.26$, (c) $\rho=0.27$, (d) $\rho=0.28$, 
(e) $\rho=0.30$, (f) $\rho=0.32$, (g) $\rho=0.34$, (h) $\rho=0.36$,
(i) $\rho=0.38$, and (j) $\rho=0.40$.
The clump phase appears for $\rho < 0.28$. 
For $0.20 \leq \rho < 0.28$, the clumps form a triangular lattice. 
For 
$0.28 \leq \rho \leq 0.44$ a stripe phase occurs. 
The stripe structures have some density modulations along their length for 
$\rho = 0.28$ in panel (d) and $\rho=0.30$ in panel (f), 
but are more uniform for the higher densities. The stripe phases
at $0.32 \leq \rho \leq 0.36$ in panels (f), (g), and (h) 
contain three rows of particles per stripe.          
}
\end{figure*}

At higher densities where the clumps contain larger
numbers of particles, the difference in the number of particles in
neighboring clumps is still only one or two particles.  As a result,
the effective polydispersity drops significantly and the system
crosses over to a limit where 
each clump has close to the same charge. 
This tendency for the clumps to have more triangular
ordering as the clumps grow in size is clearly observable
for $\rho \geq 0.2$ in Fig.~1(i,j) and 
for $0.24 \leq \rho \leq 0.27$ in Fig.~2(a-c),
where the center of 
masses of the clumps form a triangular lattice.     
In Ref.~\cite{Zapperi}, for colloidal systems 
with shorter range interactions, a larger
polydispersity in the clump sizes appeared; however, the clumps formed
an ordered triangular state whenever all of the clumps contained
six or more particles.
In the short range, purely repulsively interacting soft-sphere system
of Ref.~\cite{Glaser}, triangular ordering also appeared even for polydisperse
clumps as long as the number of particles per clump was six or greater.
Simulations of a mixed dimer and trimer clump state in a system
with a two-step repulsive interaction produced disordered states very similar
to the ones we observe \cite{Pell}.
These results indicate 
that clump states with small numbers of particles are generally disordered
independent of the exact details of the particle-particle interaction
potential, while clump states with larger numbers of particles per clump 
tend to form triangularly ordered structures.

Figures 1 and 2 indicate that the clump states form for $\rho < 0.28$. 
As the clusters grow in size with increasing density, we 
find that there is a tendency for 
clusters with certain ``magic'' numbers of particles 
to appear more often than clusters with non-magic particle numbers.
In studies of isolated clusters 
of particles with the same interaction potential,
the clusters with $n = 14$ particles were 
particularly stable \cite{Colloid}. 
A simple counting shows that clusters
of size $n=14$ appear the most often 
at the highest clump densities of $0.24 \leq \rho \leq 0.27$. 
The largest clusters of 
size $n=16$ occur at $\rho = 0.27$,  
while at $\rho = 0.28$ we observe the onset of stripe phases.

\subsection{Stripe Phases} 

In previous work on striped phases, the stripes formed disordered
labyrinth patterns
when a substrate of quenched disorder was present \cite{Martin}. 
In the absence of quenched disorder, 
aligned stripes may form depending on the rate of thermal quenching.
For rapid quench rates,
different domains of stripe orientation can arise 
leading to a partially ordered stripe state;
however, for sufficiently slow quench
rates, the lower energy aligned stripe states can form.
The orientation of the stripes is affected by the 
periodic boundary conditions and by the number of particles 
within each stripe.    

\begin{figure*}
\includegraphics[width=6.2in]{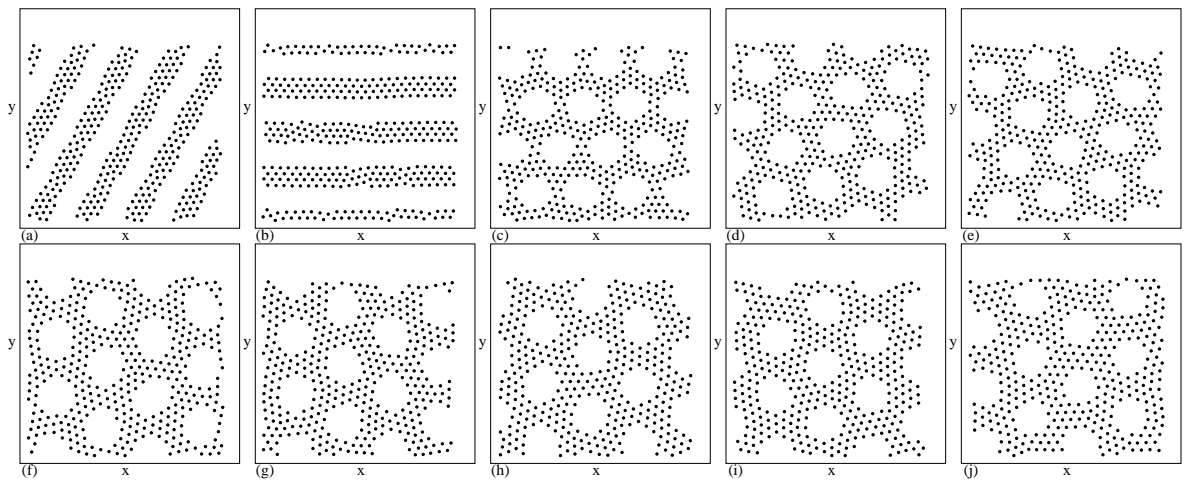}
\caption{ 
Particle positions in the entire sample after simulated annealing for 
different densities in the stripe and anticlump regimes.
The scale of the x and y axes is nearly the same in all panels.
(a) $\rho = 0.42$, (b) $\rho=0.44$, (c) $\rho=0.46$, (d) $\rho=0.47$, 
(e) $\rho=0.48$, (f) $\rho=0.50$, (g) $\rho=0.52$,
(h) $\rho=0.54$, (i) $\rho=0.56$, and (j) $\rho=0.58$.
 The stripe phase in panel (b) at $\rho = 0.44$ contains stripes that are
close to four particles wide.
The onset of the anticlump phases appears in panel (c) 
at $\rho = 0.46$.  At the low density side of the anticlump phase, the
structure is anisotropic, as can be seen by the thinning of the
necks between anticlumps along the $x$ direction in panel (c).
For densities $\rho>0.46$, the anticlump phase is no longer anisotropic.
}
\end{figure*}

At the lower-density end of the stripe regime, we observe stripes that
are composed of three rows of particles which themselves form a 
sub-triangular lattice structure within the stripe, such as at 
$\rho=0.32$ in Fig.~2(f).
For densities 
$\rho<0.32$,
the particles within the stripes are partially disordered and form
elongated quasi-clump states along the length of the stripe, creating
pinch-like structures of the type seen in Fig.~2(d) and (e) at $\rho=0.28$
and $\rho=0.30$.
For the higher densities of $0.32 \leq \rho \leq 0.34$ 
in Figs.~2(f-g), each stripe is nearly consistently three particles wide.
As the density increases, bulges in the stripe that are four particles
wide begin to form in a staggered pattern, 
as shown in Fig.~2(h,i) for $\rho=0.38$ and $\rho=0.40$. 
At $\rho = 0.38$ in Fig.~2(i), the stripes become partially disordered 
when a portion of the stripes buckle outward to form regions that are
four particles wide.
This process continues for $\rho = 0.40$ and $\rho=0.42$ 
in Fig.~2(j) and Fig.~3(a). For
$\rho = 0.44$, Fig.~3(b) shows that a large portion of the stripes are
now four particles wide.
Above $\rho=0.44$, the system transitions into the anticlump phase of the
type shown in Fig.~3(c).
For different values of 
the screening length $1/\kappa$ than we consider here,
other types of stripe phases 
containing only two rows or as many as five rows of particles 
could be realized; however, the
general evolution of phases from clump to stripe to anticlump 
as a function of increasing density should hold for any choice of parameters.   

\begin{figure}
\includegraphics[width=3.5in]{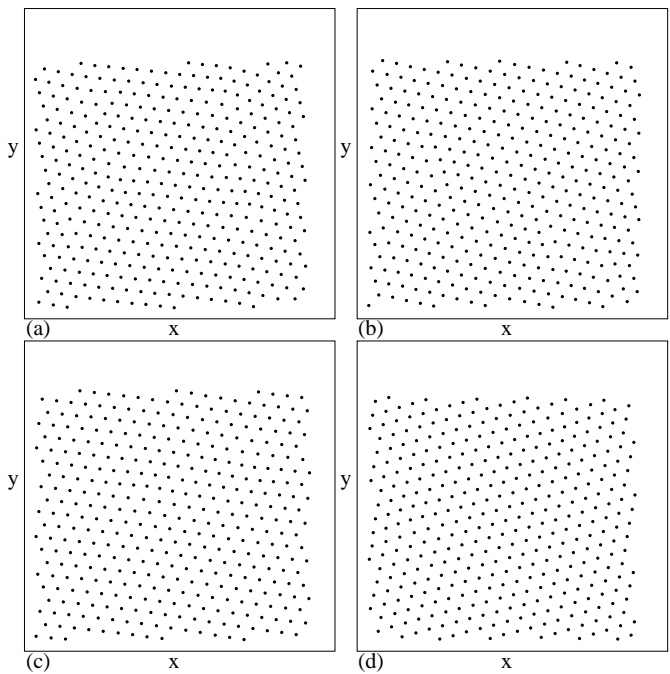}
\caption{
Particle positions in the entire sample after simulated annealing for
different densities in the uniform regime where the ordering is mostly 
triangular.
(a) $\rho=0.60$, (b) $\rho=0.64$, (c) $\rho=0.68$, and (d) $\rho=0.72$.
}
\end{figure}

\subsection{Anticlump and Uniform Phases} 

Figure~3(c) shows the 
anticlump phase, composed of a triangular lattice of voids, 
which first forms at $\rho = 0.46$. 
At the low-density onset of the anticlump phase, the anticlump state is
anisotropic and 
two of the six neck regions that 
surround each void are thinner than the other four, as seen in Fig.~3(c). 
The anticlump phases for $0.47 \leq \rho \leq 0.58$, shown
in Fig.~3(d-j), still have the triangular 
void ordering; however, the neck regions surrounding each void are now all
of equal width
so that the anisotropy found at $\rho = 0.46$ is absent. 
The particles outside of the voids attempt to form a triangular lattice 
segment that is roughly three particles wide.
For $\rho = 0.56$ and $0.58$, shown in Fig.~3(i) and Fig.~3(j),
the width of this triangular lattice grows to 
be four particles wide before a transition to a 
slightly distorted uniform 
triangular lattice phase
occurs at $\rho=0.6$.
This triangular lattice uniform state persists for all $\rho \geq 0.6$
and the triangular ordering improves with increasing density. 

\begin{figure}
\includegraphics[width=3.5in]{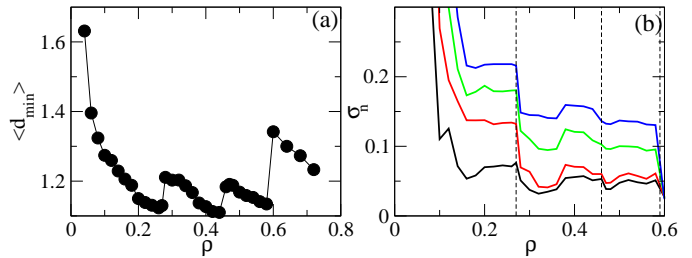}
\caption{ 
(a) Average distance to closest neighbor $\langle d_{\min}\rangle$ 
vs $\rho$. The jumps indicate transitions between the clump, stripe, anticlump,
and uniform phases.
(b) The variation $\sigma_n$ of 
the mean neighbor distance averaged over $n$ neighbors 
for $n = 3$, 4, 5, and 6 (from bottom to top). 
The dotted lines indicate the locations 
of the transitions seen in panel (a). 
Compared to $\langle d_{\min}\rangle$, two additional features
appear in $\sigma_n$.
There is a plateau in $\sigma_n$ for $0.2 \leq \rho \leq 0.28$
corresponding to the regime where the clumps form 
a triangular lattice, while for $\rho < 0.2$ the clumps are disordered. 
There is also a jump in $\sigma_n$ at $\rho = 0.38$ 
which corresponds to the density at which the stripe state changes from
having stripes that are three particles wide to having stripes that are
four particles wide.
}
\end{figure}

\subsection{Phases and Neighbor Lengths} 

To characterize the onset of the different phases, 
we compute the average closest neighbor distance
$\langle d_{\min}\rangle$ for the post-simulated annealing
configuration at each density.
To determine $\langle d_{\min}\rangle$, 
we solve the all-nearest-neighbors problem with a simple algorithm. 
The distance from each particle $i$ to its closest neighbor 
at ${\bf R}_{nn}^i$ is
$d_{\min}^i=|{\bf R}_i-{\bf R}_{nn}^i|$; 
then, $\langle d_{\min}\rangle=N^{-1}\sum_i^N d_{\min}^i$.
In Fig.~5(a) we plot $\langle d_{\min}\rangle$ as a function of 
density $\rho$.
For the lowest density clump phases, $\langle d_{\min} \rangle$ is large
due to the presence of many individual particles whose closest neighbor is
located in a different clump.
As $\rho$ increases, $\langle d_{\min}\rangle$ decreases as the particles
within each clump are compressed closer together.
The transition to the stripe phase at $\rho=0.28$ is marked by a sudden
jump in $\langle d_{\min}\rangle$.
The stripes are able to pack the available space more efficiently than the
clumps while maintaining a stripe-stripe spacing similar to the clump-clump
spacing at the end of the clump phase.  As a result, the particles within the
stripes are able to decompress and shift to a larger average spacing as they
form a partial triangular lattice within each stripe.
The neighbor spacing $\langle d_{\min}\rangle$ decreases 
with increasing $\rho$ throughout the stripe phase as the particles within the
stripe compress in order to maintain the stripe structure.
The transition into the anticlump state at $\rho=0.46$ is again accompanied by
a sudden increase
of $\langle d_{\min}\rangle$ as the particles expand in order to
take advantage of the additional area available to them in the anticlump
structure.
Within the anticlump phase, $\langle d_{\min}\rangle$ decreases with
increasing $\rho$ as the particles are compressed into 
the diminishing space between the voids.
The  transition into the uniform phase appears as a large jump in
$\langle d_{\min}\rangle$ at $\rho = 0.6$. 
This jump is significantly larger than the other jumps 
in $\langle d_{\min}\rangle$ since the particles are now able to fill the
entire system evenly.  As $\rho$ continues to increase within the
uniform phase, $\langle d_{\min}\rangle$ decreases as the lattice constant
of the triangular lattice shrinks.
The sharpness of the transitions in $\langle d_{\min}\rangle$ suggest that
these transitions may be first order in nature.     
In addition to computing the nearest neighbor distance $d_{\min}$ for
each particle, we can determine the distance $d_i^{n_j}$,
$n_j \in (1...n_n^i)$ between a particle and
each of its $n_n^i$ neighbors as identified by a Voronoi construction
\cite{Fortune}.
The distribution $P(d_i^{n_j})$ of $d_{i}^{n_j}$ at a particular value of $\rho$
is bidisperse for $\rho < 0.6$ due to the presence of two length
scales: the short spacing between neighbors that are within the same
clump, stripe, or on the same side of a void, and the longer spacing between
neighbors that are in adjacent clumps or stripes or are on opposite sides of
a void. 

We compute the variation in particle spacing throughout the system at different
values of $\rho$ using a measure $\sigma_n$ constructed in the following
way: For each particle, we determine the mean distance $d_n^i$ from that
particle to its $n$ closest neighbors: 
$d_n^i=n^{-1}\sum d_i^{n_j}$ for $n_j$ ranging over the $n$ closest neighbors
only.
We then determine the mean value $\mu_n$ of $d_n$ averaged over all
particles, $\mu_n=N^{-1}\sum_i^{N}d_n^i$, and finally obtain the
standard deviation $\sigma_n$ of this quantity,
$\sigma_n=\sqrt{(N-1)^{-1}\sum_i^{N}(d_n^i-\mu_n)^2}$.
In Fig.~5(b) we plot $\sigma_n$ as a function of $\rho$ for the number of
closest neighbors $n=3$, 4, 5, and 6.  At small values of $\rho$ when 
the clumps contain only one or two particles, the measure $\sigma_n$ performs
poorly for $n>2$, but for larger clumps and in all the remaining phases,
the measure is well defined for $n<7$.
The dashed lines in Fig.~5(b) indicate the locations of the transitions
found in Fig.~5(a).
As $n$ increases, the magnitude of the variance $\sigma_n$ also
increases for any given density $\rho$, which is expected since a larger
number of neighbors are being averaged.
Several of the features in $\sigma_n$ correlate with the transitions
found in $\langle d_{\min}\rangle$ in Fig.~5(a); however, $\sigma_n$ 
captures at least two features which do not appear
clearly in Fig.~5(a). 
The first is that for $n = 4$, 5, and 6, $\sigma_n$ 
rapidly decreases with increasing $\rho$ up to $\rho = 0.2$ and then 
reaches a constant plateau which persists until the transition into the
stripe state at $\rho =0.28$. 
The onset of the plateau
at $\rho = 0.2$ is correlated with the density at which
the clumps start to form a triangular lattice, while
for $\rho < 0.2$ the clump structure is much more disordered. 
This may indicate that the 
charge polydispersity effect becomes small enough 
at $\rho=0.2$ that a transition to a triangular clump lattice can occur.
The second distinctive feature is the increase in $\sigma_n$
that occurs near $\rho = 0.38$ within the stripe
regime. This feature is correlated with the density at which the
stripes change from being an average of three particles wide to being
an average of four particles wide.
At the onset of the uniform phase at $\rho=0.6$, $\sigma_n$ drops 
and collapses so that all $\sigma_n$ take the same value even for different
choices of $n$.
The $\sigma_n$ measure indicates that in addition 
to the primary clump-stripe-anticlump phases, there can be sub-phases
associated with changes in the average stripe width or
the onset of ordering of the clumps. 

\begin{figure}
\includegraphics[width=3.5in]{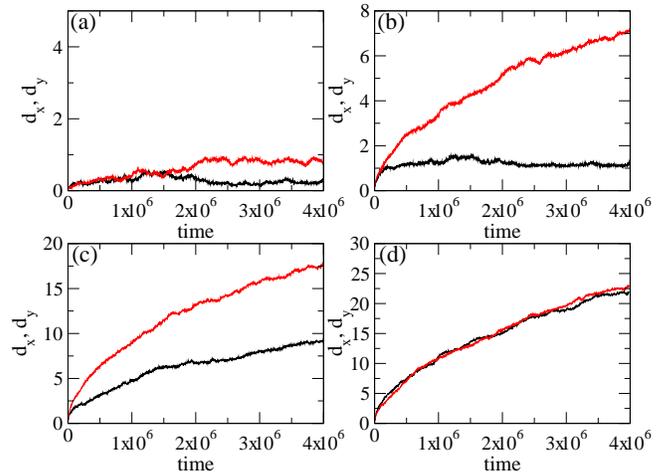}
\caption{
The cumulative particle displacements  $d_{x}$ (dark lower line) and $d_{y}$ 
(light upper line)
vs time at different temperatures
for the stripe sample at $\rho=0.36$ in which the stripes are aligned
along the $y$ direction. 
(a) At 
$T=0.031$
both $d_x$ and $d_y$ are bounded.
(b) At 
$T=0.78$,
the increasing $d_y$ indicates that diffusion is occurring 
in the $y$-direction along the length of the stripes, but 
the flat $d_x$ indicates that there is no diffusion transverse to the stripes.
(c) At 
$T=2.0$,
the stripe structure is still present and the diffusion
is anisotropic with $d_y>d_x$.  There is now finite diffusion transverse
to the stripes.
(d) At 
$T=2.53$,
the stripe structure is melted 
and the diffusion is isotropic.    
}
\end{figure}

\begin{figure}
\includegraphics[width=3.5in]{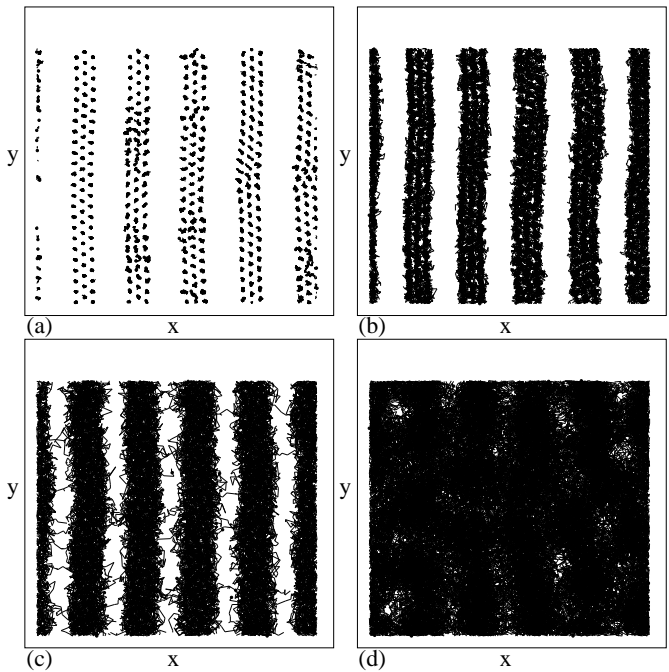}
\caption{ 
The particle positions (dots) and trajectories (lines) for a fixed period 
of time at different temperatures for the stripe phase at $\rho = 0.36$ 
highlighting the different diffusive regimes in Fig.~6.
(a) At 
$T=0.031$,
the particles in the stripe are frozen. 
(b) At 
$T=0.78$,
the particles
can diffuse along the stripes but there is no diffusion between stripes.
(c) At 
$T=2.0$,
the stripe structure is still present 
and motion occurs both along the stripes and across the stripes. 
It remains easier for the particles to diffuse along the stripes than 
across the stripes, producing an anisotropic diffusion signature.
(d) At 
$T=2.53$,
the stripe structure is destroyed and the particle motion
is isotropic.
}
\end{figure}

\section{Temperature Induced Disorder}

\subsection{Different Liquid Regimes in the Stripe Phase} 

We first consider melting transitions in the stripe state
at $\rho = 0.36$, where the stripes are aligned in the 
$y$-direction as shown in Fig.~2(h). 
To examine the effect of temperature, 
we calculate the particle displacements $d_x$ and $d_y$ as a 
function of time for this phase at different 
temperatures for both the $x$ and $y$ directions.
\begin{equation}
d_{x,y}(t) = (1/N)\sum^{N}_{i}|R_{x,y}(t) -R_{x,y}(t_0)|.      
\end{equation}
In an isotropic liquid, we expect $d_x=d_y$.
In Fig.~6 we plot $d_{x}$ and $d_{y}$ at temperatures
$T = 0.031$, 0.78, 2.0, and $2.53$. 
Figure 6(a) shows that at 
$T=0.031$
the displacements are bounded
at long times and that $d_{x}$ and $d_{y}$ 
have approximately the same value. In Fig.~7(a),
the particle trajectories during a fixed time at 
$T=0.031$
indicate that
the particles undergo small displacements due to the thermal 
fluctuations 
but that both the stripe ordering and the crystalline
ordering within the stripes are preserved, 
showing that the system is in a frozen state.
We note that for much longer times than those plotted in Fig.~6(a), 
the values of $d_{x,y}$ do not increase but remain bounded at a nearly
constant level.
At 
$T=0.78$,
plotted in Fig.~6(b), $d_{y}$ 
grows without bound while $d_{x}$ saturates
and is bounded. Fig.~7(b) shows that   
for 
$T=0.78$
the particles readily diffuse
along the stripes 
but there is no diffusion between the stripes. 
The fragmentary triangular particle ordering within the stripes is lost and 
an interesting state appears 
that acts like
a liquid in one direction but like a solid in the other 
direction, which is characteristic of a smectic phase.   
For 
$T=2.0$,
Fig.~6(c) indicates that both $d_{y}$ and $d_{x}$ 
increase with time; however, $d_{y}$ rises more
rapidly indicating that the diffusion is anisotropic. 
In Fig.~7(c) the particle trajectories at 
$T=2.0$
show  
that there is a large amount of motion along the stripes 
accompanied by 
a smaller number of places where the particles have hopped from 
one stripe to the next.  The overall stripe structure remains intact. 
For 
$T=2.53$,
plotted in Fig.~6(d), $d_{x}$ and $d_{y}$ 
both increase at the same rate, indicating isotropic diffusion.
Figure~7(d) shows that the stripe structure is destroyed 
at 
$T=2.53$
and the particle motion  
is uniform. 
Although the liquid state in Fig.~7(d) 
is isotropic, we observe
short-lived clumps that do not occur in an isotropic liquid
of purely repulsive particles \cite{Goree}.  

\begin{figure}
\includegraphics[width=3.5in]{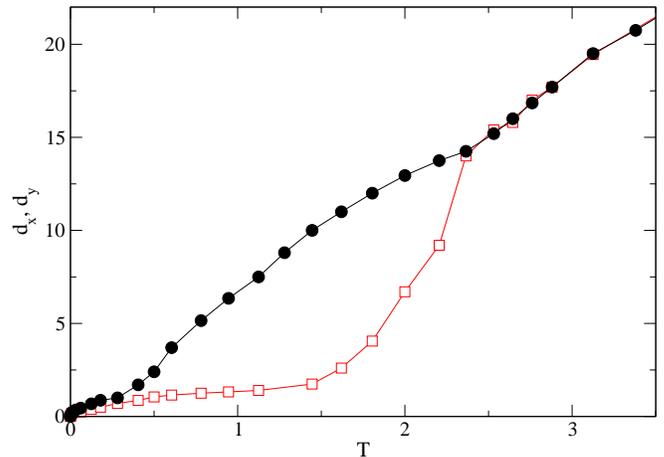}
\caption{ 
The value of the cumulative particle displacements 
$d_x$ (open squares) and $d_y$ (filled circles)
after $2\times 10^6$ simulation time steps as a function of 
temperature $T$ for the 
stripe system at $\rho = 0.36$. 
At low temperatures, there is a frozen regime with $d_x=d_y$; at
intermediate temperatures, an anisotropic liquid appears with 
$d_y>d_x$; and at high temperatures an isotropic liquid forms with
$d_x=d_y$.
}
\end{figure}

To better characterize the onset of the different 
diffusive regimes we plot the value of 
$d_x$ and $d_y$ after $2\times 10^6$ 
simulation time steps as a function of temperature $T$ in Fig.~8. 
For 
$T<0.32$,
the particles within the stripe remain
frozen and the diffusion is isotropic. 
For 
$0.32 \leq T \leq 1.53$,
the particles can flow in a liquidlike fashion
along the length of the stripes but there is no diffusion transverse
to the stripes.
Within this phase, if we measure $d_{x,y}$ 
for increasingly long times, the value of $d_{y}$ 
continues to increase but $d_{x}$ remains at a plateau value corresponding
to the width of the stripe, which is the
maximum distance the particles can diffuse in the $x$ direction in 
this regime.
For 
$1.53 < T \leq 2.37$,
the stripe structure is still present although
there is now some diffusion transverse to the stripes. 
The diffusion remains anisotropic with $d_{y} > d_{x}$.
Finally for  
$T>2.37$,
the diffusion becomes isotropic when the stripe
structure is destroyed.

\begin{figure}
\includegraphics[width=3.5in]{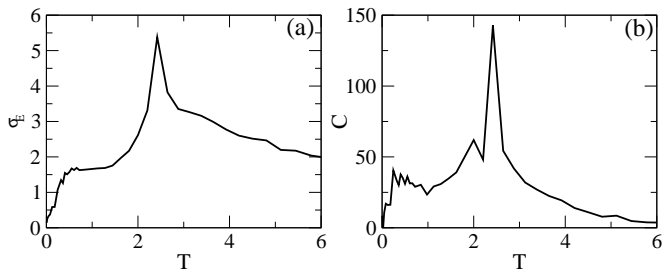}
\caption{ 
(a) The standard deviation of the energy 
$\sigma_E$ vs $T$ for the stripe phase at $\rho = 0.36$. 
The large peak at $T=2.4$ 
corresponds to the melting of the entire stripe structure. 
The smaller knee structure near $T=0.5$ corresponds to the
onset of melting of the particles within the stripes. (b)
The specific heat $C$ vs $T$ for the same system shows 
a large peak at the stripe melting temperature and some features near
the intra-stripe disordering transition.
}
\end{figure}

\begin{figure}
\includegraphics[width=3.5in]{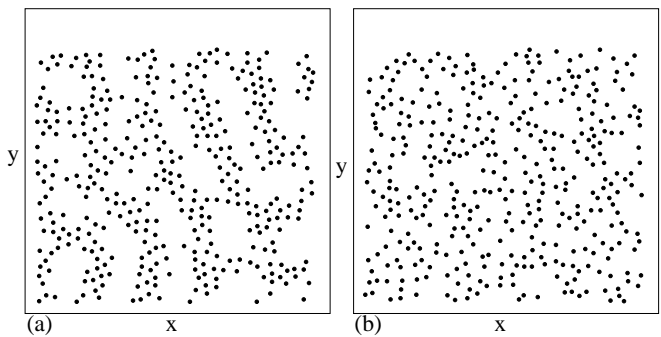}
\caption{
The particle positions at one instant in time for the stripe phase
at $\rho=0.36$. 
The scale of the $x$ and $y$ axes changes in each panel.
(a) 
$T=2.53$ showing transient clump structures.
(b)
$T=6.13$ showing that the transient clumping is greatly reduced.
}
\end{figure}

To further characterize the stripe melting, in Fig.~9(a)  we plot 
$\sigma_E$, the standard deviation of the energy fluctuations, as a 
function of temperature.
We obtain this measurement by collecting a time series of the 
total particle-particle interaction energy $E$
during $5\times 10^6$ simulation time steps while holding the system at
a fixed temperature.
We also compute the average energy $\bar E$ at each temperature
and use this to construct the specific heat, $C=d\bar E/dT$, plotted in Fig.~9(b).
Both the fluctuations $\sigma_E$ and the specific heat $C$ show a pronounced
peak at $T=2.4$ when the stripe structure melts into an isotropic
liquid state.
A knee structure appears in both measures near $T=0.3$ at the onset of
the intra-stripe liquid phase.  The magnitude of the energy fluctuations
$\sigma_E$ remains nearly constant for $0.3 < T < 1.4$, inside the 
intra-stripe liquid.
Both $\sigma_E$ and $C$ gradually decrease with increasing temperature above the
stripe melting temperature.
For purely repulsive particle assemblies, the magnitude of the 
energy fluctuations falls off very rapidly above the melting
temperature.
The slow decrease in $\sigma_E$ for $T>2.4$ in the liquid state of the
stripe system
reflects the presence of short-lived clumps within the liquid which become
smaller and even shorter-lived as $T$ increases.
In Fig.~10(a) we show a snapshot of the particle positions in the 
isotropic liquid phase at 
$T=2.53$, showing that the particle positions remain highly inhomogeneous
on short time scales and form clump-like structures that change
rapidly with time.  For contrast, in Fig.~10(b) we show a similar snapshot
deeper into the isotropic liquid phase at 
$T=6.13$, indicating that the transient clumping effect is significantly
reduced at higher temperatures.
In Fig.~9(b), $C$ has the same trends as $\sigma_E$; 
however, the data is generally noisier. 
At the onset of the intra-stripe liquid near $T=0.3$, a smaller peak-like
structure appears in $C$, indicating an excess of entropy.

An open question is whether the different regimes 
observed in the stripe state as a function of temperature
are true phase transitions
or only crossovers.
The specific heat data indicates that 
the melting transition into the fully isotropic liquid state is most likely a
phase transition associated with a change in symmetry. 
Simulations of similar stripe and pattern forming systems
generally find that the transition from a stripe state to an isotropic 
state is first order \cite{Pell}, or 
in some cases is a dislocation unbinding transition \cite{Singer}. 
Hysteresis measurements
suggest that the transition in our system is first order; 
however, our simulation size 
is not large enough to rule out a dislocation-mediated transition 
in the stripe phase.  
The much weaker peak in $C$ at the onset to the intra-stripe liquid phase 
suggests that the onset of this regime may be a crossover.  
We note that in most of our realizations of stripe phases,
there are usually some topological defects present 
in the triangular ordering of the particles within the stripe. 
These defects may permit low energy particle motions
that could obscure a true peak in the specific heat
at the onset of the intra-stripe liquid. 
In general, for a one-dimensional system there are no
finite temperature phase transitions; however,
in the stripe system, the ordering within the stripe is 
two-dimensional in nature and the particles in neighboring stripes are
coupled to one another.  Thus, a truly frozen state could occur in our
quasi-one-dimensional stripes at finite temperature
if the particles within the stripes were completely ordered. 
The onset of the anisotropic diffusive regime is most likely a crossover since 
individual particles within a stripe can be 
thermally activated out of the stripe. Due to the
structure of the stripes, there should be a well-defined 
characteristic energy required for 
these hops to occur.  When the temperature corresponding to this
energy is reached,  
the inter-stripe diffusion rapidly increases.      

\begin{figure}
\includegraphics[width=3.5in]{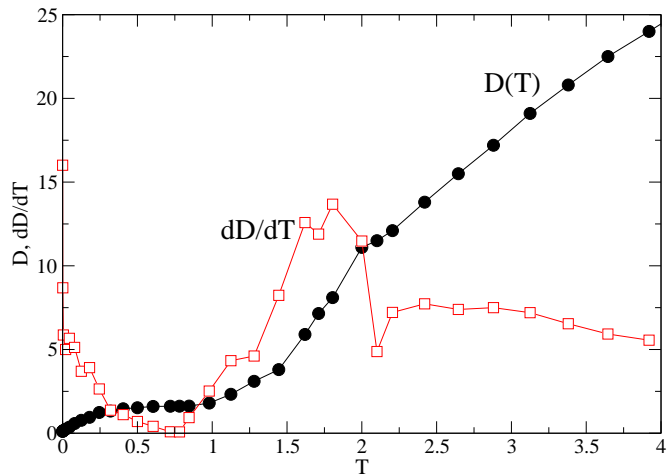}
\caption{
The cumulative particle displacements 
$D$ after $2\times10^{6}$ simulation time steps vs $T$ for the 
clump phase at $\rho = 0.26$. 
The displacements increase rapidly once the clump structure is
thermally destroyed. 
Unlike the stripe phase, the diffusion in the clump phase is isotropic.    
Also plotted is $dD/dT$ vs $T$,
showing 
a peak at the clump melting transition 
near $T=2.0$ along with a small
rounded peak 
near $T=0.25$, which corresponds to the 
melting of the particles within each clump.
}
\end{figure}

\subsection{Different Liquid Regimes Within the Clump and Anticlump Phases} 

We next consider the same set of measurements for the 
clump phase at $\rho = 0.26$. 
In Fig.~11 we plot $D=d_x=d_y$, the isotropic particle displacements
after $2 \times 10^6$ simulation time steps, as a function of temperature
in the clump phase.
The diffusion is isotropic and Fig.~11 indicates that the displacements rapidly
increase just below $T = 2.0$ with a cusp type feature 
appearing at $T = 2.0$. 
We also plot the derivative $dD/dT$, which passes through a peak just
below $T=2.0$ at the point when the clumps begin to break apart.
There is also a small peak in $dD/dT$ near 
$T=0.25$ associated with the melting of the particles within each clump.
If we perform the measurement over a longer period of time,
the values of $D$ for 
$T<1.125$
remain unchanged
while the values of $D$ for 
$T \geq 1.125$
increase.

\begin{figure}
\includegraphics[width=3.5in]{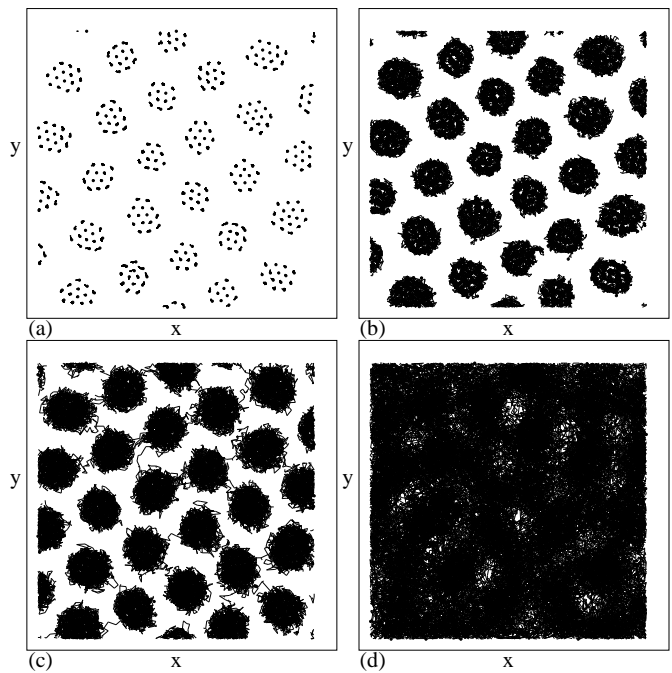}
\caption{
The particle positions (dots) and trajectories (lines) 
for a fixed period of time at different temperatures for the clump phase
at $\rho = 0.26$. 
(a) 
At $T=0.031$ the clumps form a lattice and the particles within the clumps
show little motion. 
(b) 
At $T=0.78$, the clumps still form a lattice but the particles are diffusing
within each clump.
(c) At 
$T=1.28$ the clump structure persists but interclump diffusion is now
occurring in addition to intraclump diffusion.
(d) 
At $T=3.125$ the clump lattice structure is destroyed
and diffusion occurs freely throughout the system.
}
\end{figure}

In Fig.~12(a) we illustrate the particle positions and trajectories 
in the clump state at $\rho = 0.26$ for 
$T=0.031$.
Here the clump structure is ordered 
and the particles within each clump undergo little motion.
At $T=0.78$, shown in Fig.~12(b),
the clump structure is still present but the particles within each
clump are highly disordered and diffuse rapidly within the clumps.  
This intra-clump liquid state has many similarities to 
the intermediate stripe liquid state.
The overall structure of the system is ordered but subregions 
of the system are liquefied. In the
stripe case, it was possible for the particles to diffuse 
along the full length of the stripes so that diffusion in this direction
was unbounded.  For the clump state, the diffusion in the intra-clump
liquid always remains bounded by the finite clump size.
In the stripe system the 
sharpness of the transition from the frozen stripe crystal 
to the intrastripe liquid is controlled by the amount of ordering
within the stripe phase. 
In the clump system, a similar effect occurs. 
It was shown in previous work
that there are certain magic sizes of individual clumps that produce
highly ordered configurations of the particles within the clump \cite{Colloid}.
For example, a clump containing 14 particles can form a stable 
segment of a triangular lattice which is free of internal topological
defects.
Additionally, studies of
small numbers of particles in individual traps also indicate 
that the onset of rotational and translational
disordering is affected by the number of particles in the trap 
and the degree of order in their arrangement 
\cite{Lin,Drocco,Leiderer,Peeters2}. 
For our clump system, this implies that
if all the clumps had the same number of particles and each clump was
also well ordered internally, such as for $n=14$ particles per clump,
a well defined transition from a frozen clump phase to 
an intra-clump liquid phase should occur. 
In our system, not all the clumps have the
same number of particles and some clumps
contain particle numbers which produce less stable internal
clump structures; therefore, our sample shows
disorder at lower temperatures which smears out the transition
to the intra-clump liquid.      

In Fig.~12(c) we plot the particle trajectories at 
$T=1.28$ in the clump state.
The clump structures persist and the particles within each clump have
liquefied; however, there is now some diffusion of particles from clump
to clump.
This phase is similar to the stripe phase which showed
anisotropic diffusion.  In the clump
system, the diffusion always remains isotropic. 
If we follow the trajectory of a single particle, we observe that it
remains trapped within a single clump for a period of time, diffusing within
that clump, before escaping from the clump and undergoing much more rapid
diffusion until it is trapped by another clump.
This type of motion is very similar to the caging dynamics
seen in certain glassy systems where particles 
can locally diffuse in some region
with an occasional long hop to another region where the particle again
undergoes localized diffusion. 
At very short times, the behavior appears purely diffusive when the
particles have not had enough time to experience the 
confining boundaries of the clump.
At intermediate times, the behavior is subdiffusive due to the clump 
confinement, while at very long times, the behavior is diffusive since
clump-clump diffusion can occur at this temperature.
A similar behavior occurs for the stripe phases 
for diffusion in the direction perpendicular
to the stripes. 
This behavior suggests that these regions may exhibit 
some glassy type behaviors, a topic which we will
explore in a separate work.

In Fig.~12(d) for 
$T=3.125$,
the clump lattice is destroyed and the system enters
an  isotropic liquid state.
Just as in the stripe case, the liquid phase at temperatures 
not too far above the clump melting region
shows short-lived density inhomogeneities which decrease in size 
and lifetime as      
the temperature increases. 

\begin{figure}
\includegraphics[width=3.5in]{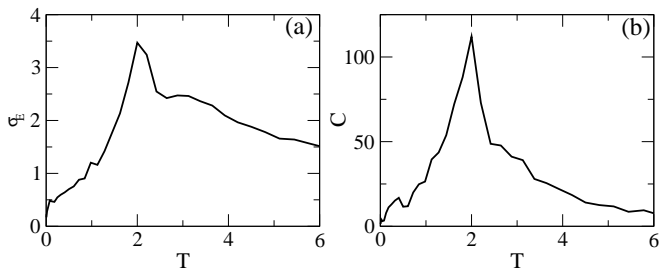}
\caption{(a) The standard deviation of the energy 
$\sigma_E$ vs $T$ for the clump phase at $\rho = 0.26$.
There is a peak at $T=2.0$ at the clump melting transition. 
(b) The specific heat $C$ vs $T$ for the same system
shows a similar peak at the
clump melting temperature. 
In contrast to the stripe phase, 
there is no discernible feature 
in either quantity at the intra-clump melting transition.
}
\end{figure}

In Fig.~13(a) we plot $\sigma_E$ versus $T$ for the clump state at
$\rho=0.26$
and in Fig.~12(b) we plot the corresponding specific heat $C$ vs $T$. 
Both quantities show a peak at $T=2.0$
where the clump structure breaks apart,
followed by a slow decay at higher temperatures in the isotropic liquid regime. 
Neither quantity shows any noticeable features at the 
frozen clump solid to intra-clump liquid melting
which is likely due to the 
fact that the clumps contain unequal numbers of  particles.

\begin{figure}
\includegraphics[width=3.5in]{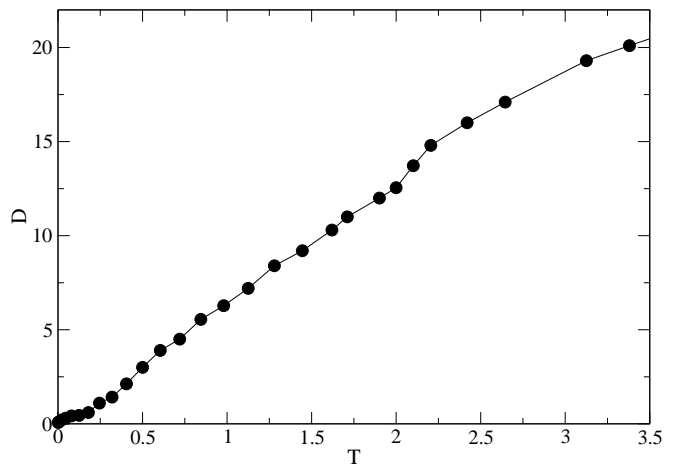}
\caption{
The cumulative particle displacements $D$ after $2 \times 10^6$
simulation time steps vs $T$ for the anticlump phase at $\rho = 0.54$. 
Here the onset of diffusion occurs
at 
$T=0.28$ corresponding to the disordering of the particles within the
non-void regions.  This is followed by a second change in the
behavior of $D$ at
$T = 2.0$, which corresponds to the melting 
or destruction of the void lattice.
}
\end{figure}

\begin{figure}
\includegraphics[width=3.5in]{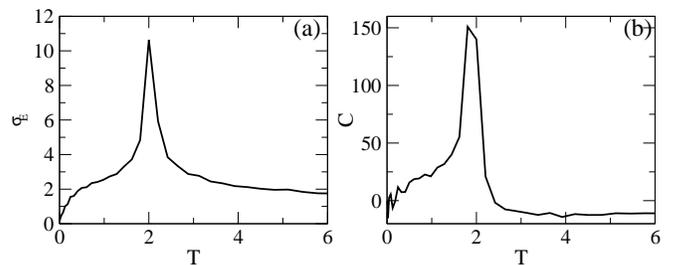}
\caption{
The standard deviation of the energy $\sigma_E$ vs $T$ for the 
anticlump phase at $\rho = 0.54$.
There is a peak at $T = 2.0$ where the void lattice melts. 
There is also some indication of a change of slope
near 
$T=0.28$.
(b) The specific heat $C$ vs $T$ for the same system
shows a similar peak at $T = 2.0$. The specific heat falls off much more
rapidly above the melting transition than in the clump or stripe phases
due to the fact that the isotropic liquid for the anticlump phase does not
possess the same type of density inhomogeneities found for the clump and
stripe phases.
}
\end{figure}

We next examine the effect of temperature on the anticlump phase 
at $\rho = 0.54$. 
In Fig.~14(a) we show the cumulative displacements 
$D=d_x=d_y$ after $2\times 10^6$ time steps versus $T$.
The displacements are isotropic and show a significant increase 
near 
$T=0.28$ followed by a 
smaller change 
near $T = 2.0$. The triangular void crystal 
remains intact up to $T = 2.0$, but diffusion through the
entire sample begins at significantly lower temperatures. 
In Fig.~15(a) we plot $\sigma_E$ for the anticlump phase
at $\rho=0.54$.
There is a very sharp peak centered at $T = 2.0$ 
corresponding to the melting transition of the voids.
There also is some indication of a change in slope near 
$T=0.28$.
In Fig.~15(b) we show the corresponding specific heat $C$ versus $T$.
There is a peak in $C$ at $T = 2.0$ but there is no particular feature at 
$T=0.28$.
The specific heat falls off rapidly for $T=2.0$
due to the fact that the liquid phase does not exhibit the
same large short-lived density inhomogeneities 
found above melting in the stripe and clump phases.

\begin{figure}
\includegraphics[width=3.5in]{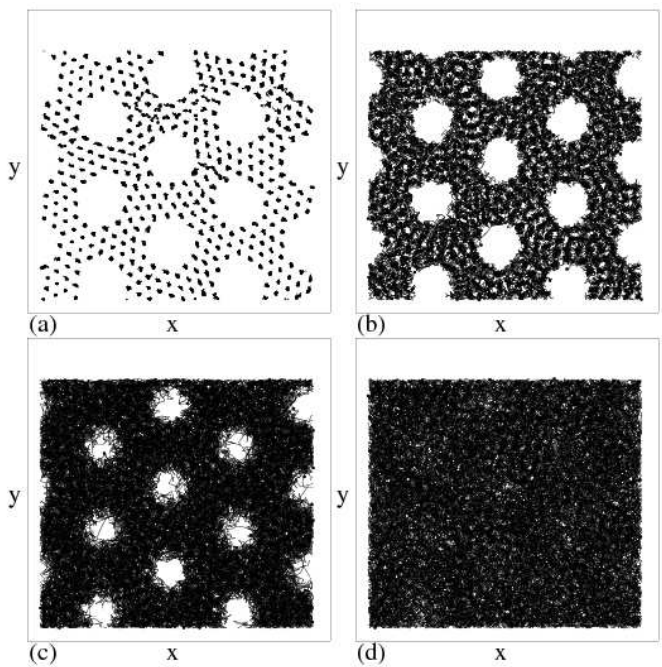}
\caption{
The particle positions (dots) and trajectories (lines) for a fixed period
of time at different temperatures for the anticlump phase at $\rho = 0.54$. 
(a) 
At $T=0.125$
the particles undergo little motion and the state is frozen.
(b) 
At $T=0.72$ the void lattice persists 
but the particles have liquefied
and diffuse throughout the non-void region. 
(c) 
At $T=1.71$ the behavior is similar to that for
$T=0.72$ 
but with even more motion in the non-void regions. 
(d) 
At $T=2.42$
the void lattice is completely destroyed and 
diffusion occurs throughout the sample.  
}
\end{figure}

In Fig.~16 we illustrate the particle trajectories at different temperatures
in the anti-clump phase at $\rho=0.54$.
At $T=0.125$, shown in Fig.~16(a), 
the particles undergo little motion indicating that
the system is frozen. 
In Fig.~16(b) at 
$T=0.72$,
the triangular void structure remains present but there
is now particle motion through the non-void regions, 
indicating that particles are able to diffuse through the entire 
system. 
In Fig.~16(c) at 
$T=1.71$,
the void structure persists but some particles can now move
into the voids for short periods of time. 
The behavior for 
$0.28 < T < 2.0$
has the same features
found in the intermediate liquid phase 
for the stripe state where the individual particles have liquefied but
the overall structure of the state remains solid. 
In Fig.~16(d) at 
$T=2.42$,
the void structure has melted and
diffusion can occur anywhere in the system. 

\begin{figure}
\includegraphics[width=3.5in]{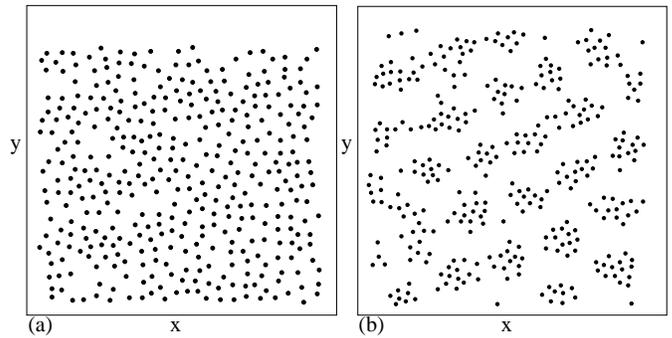}
\caption{
The particle positions at one instant in time at 
$T=2.42$.
(a) The anticlump phase at $\rho=0.54$.
(b) The clump phase at $\rho=0.26$
}
\end{figure}

In Fig.~17(a) we plot the instantaneous particle positions in the 
anticlump phase for $\rho=0.54$ at 
$T=2.42$ and in Fig.~17(b) we show the corresponding particle positions
at the same temperature in the clump phase at $\rho=0.26$.
Just as in the stripe phase, the clump phase shows strong density
inhomogeneities in the liquid state that decrease in size and lifetime
with increasing temperature.  In the anticlump phase, the liquid state
found above melting is much more homogeneous and as a result the
fluctuations in the energy are much smaller for the anticlump system
above melting than for the stripe or clump phases.  We also find a
uniform density liquid for the higher density uniform regime above its
melting temperature (not shown).
    
These results indicate that the clump, stripe, and
anticlump phase all exhibit an intermediate fluid phase, 
where the internal ordering of individual particles is lost
and selective diffusion can occur while
the overall structure of the clump, stripe, or anticlump pattern remains
intact.       

\section{Discussion}
One of our main findings is that the competing long-range repulsive and
short-range attractive interactions produce a system that can exhibit
both solid and liquid behavior simultaneously.  The liquid-like behavior
occurs on the short length scale of the spacing between individual
particles.  Motion on this small length scale initiates at temperatures
well below the temperature at which the large scale patterns formed by
the particles break apart.  Future studies of the impact of the coexisting
liquid and solid behavior could include determining how the existence of
the two length scales affects measurable time scales.  For example, the
short-scale liquid could be probed with short-time measurements while
the larger-scale solid state could be accessed using longer time
measurements.  The liquid-like nature of the system on short length scales
might also play a role in permitting the system to adapt itself easily to
a substrate or to an applied external field.  For example, the ordering of
a large length scale might be disrupted by quenched disorder in a system that
had no short length scale degrees of freedom, whereas in our competing
interaction system, the effect of the disorder could be compensated on the
short length scale, permitting the large length scale ordering to be maintained
intact.  In essence, the existence of the short length scale degrees of
freedom permits a screening of the disorder on the larger length scale.
It would also be interesting to see which of the phases we observe, such
as clump, stripe, or anticlump, is more adaptable to quenched disorder
in this sense. 
Our system may have interesting connections to other types of 
disordered systems which exhibit liquid-like behaviors on some length and
time scales but solid-like behaviors on other length and time scales.
For example, glassy systems have properties that are in some sense inverted
from what we observe, since glasses have liquid-like behavior on long length
and time scales but solid-like behavior on short time and length scales.
The time and length scale separation produced by competing interactions has
clear potential for controlling novel functionality in materials, particularly
those materials which contain additional degrees of freedom such as 
electronic, magnetic, superconducting, optical, biological, and so forth.
The competing interactions can achieve a low-dimensional confinement of
active regions, such as our short length scale liquid, embedded in a 
self-consistent supporting matrix, such as our larger length scale solid.

\section{Summary}
In summary, we have analyzed the different types of 
phases that can occur in a two-dimensional system with a
competing long-range repulsion
and a short range attraction as a function of density and temperature.
At zero temperature and low densities, the system forms 
a clump phase. The particles cluster together in clumps and,
due to the long range repulsion, the clumps move as far apart from 
each other as possible and attempt to form a macroscopic triangular lattice. 
When there are few particles in each clump, the overall structure of the
clump lattice is disordered due to the large charge polydispersity of the
small clumps.
As the number of particles in each clump increases, this 
polydispersity effect is reduced 
and a triangular clump lattice appears.
In addition to the triangular ordering of the clumps, there is also a tendency
for the particles within each clump to form ordered or 
partially ordered structures. 
For higher densities, there is a sharp transition into
a stripe state.  The particles within each 
stripe tend to have triangular or partial triangular ordering.
At still higher densities, a transition into 
an anticlump phase containing a triangular lattice of voids 
occurs, and at even higher
densities there is a transition into a uniform phase 
with partial triangular ordering.
We introduced several new measures to characterize the onset of 
the different phases as a function of density, including the
calculation of the average distance to the closest neighbor. 

As a function of increasing temperature, 
we find that the stripe phase exhibits
clear multiple-step melting.
A frozen phase with little or no diffusion occurs at low temperature.
With increasing temperature the system melts into an intra-stripe liquid,
where particles diffuse along the length of the stripes but do not diffuse
between stripes and the overall stripe ordering remains intact.
At higher temperatures, there is an onset of stripe-to-stripe particle
motion leading to anisotropic diffusion, and the stripe structure continues
to persist.  Finally, there is a transition
to an isotropic liquid where the stripe structure is destroyed and the 
diffusion becomes isotropic.
The onset of these different regimes produces signatures in the
specific heat and in the energy fluctuations. 
We argue that the existence of an intermediate intra-stripe liquid
indicates that 
excess entropy can be put into the motion of particles 
along the stripes rather than into destruction of the overall stripe structure.
The presence of excess entropy is manifested as features in the 
specific heat and energy fluctuations.         
We show that the clump and anticlump phases exhibit 
similar intermediate liquid regimes.  In the clump phase,
the particles remain confined within the clumps 
but are liquidlike and mobile inside the clumps, 
while the clump lattice remains solid. For the
anticlump phase, the particles diffuse throughout the entire lattice 
while the void structure remains crystalline.  

This work was carried out under the auspices of the 
NNSA of the 
U.S. DoE
at 
LANL
under Contract No.
DE-AC52-06NA25396.

\end{document}